# Enhanced Organic Solar Cells Efficiency through Electronic and Electro-optic Effects Resulting from Charge Transfers in Polymer Hole Transport Blends

Calvyn T. Howells,♣[ab] Khalid Marbou,♣[ab] Haeri Kim,‡[c] Kwang Jin Lee,‡[d] Benoît Heinrich,[e] Sang Jun Kim,[f] Aiko Nakao,[g] Tetsua Aoyama,[g] Seiichi Furukawa,[h] Ju-Hyung Kim,[dhi] Eunsun Kim,[d] Fabrice Mathevet,[j] Stéphane Mery,[e] Ifor D.W. Samuel,[b] Amal Al Ghaferi,[a] Marcus S. Dahlem,[a] Masanobu Uchiyama,[gk] Sang Youl Kim,[fl] Jeong Weon Wu,[d] Jean-Charles Ribierre,[h] Chihaya Adachi,[h] Dong-Wook Kim,[d] Pascal André*[bdg]

- a- Masdar Institute of Science and Technology, Abu Dhabi, UAE
- b- School of Physics and Astronomy, University of St Andrews, SUPA, St Andrews, UK
- c- Clean Energy Research Center, Korea Institute of Science and Technology (KIST), Seoul, Republic of Korea
- d- Department of Physics and CNRS-Ewha International Research Center (CERC), Ewha W. University, Seoul, Republic of Korea
- e- Institut de Physique et Chimie des Matériaux de Strasbourg (IPCMS), CNRS-Université de Strasbourg (UMR 7504), Strasbourg, France
- f- Ellipso Technology (Co. Ltd.), Suwon, Republic of Korea
- g- RIKEN, Wako, Saitama, Japan. Email: pjpandre@riken.jp
- h- Kyushu University, Center for Organic Photonics and Electronics Research (OPERA), Fukuoka, Japan
- i- Department of Chemical Engineering, Pukyong National University, Busan Republic of Korea
- j- Institut Parisien de Chimie Moléculaire, Chimie des Polymères, CNRS-UMR 8232, Université Pierre and Marie Curie, Paris, France
- k- Graduate School of Pharmaceutical Sciences, University of Tokyo, Tokyo, Japan
- l- Department of Physics, Ajou University, Suwon, Republic of Korea

♣ and ‡ These authors contributed equally to this work.

We demonstrate that blending fluorinated molecules in PEDOT:PSS hole transport layers (HTL) induces charge transfers which impact on both charge extraction and photogeneration within organic photovoltaic (OPV) devices. OPVs fabricated with modified HTL and two photoactive polymer blends led systematically to power conversion efficiencies (*PCE*) increases, with PTB7:PC$_{70}$BM blend exhibiting *PCE* of ~ 8.3 %, i.e. ~ 15 % increase compared to pristine HTL devices. A reduced device-to-device characteristics variations was also noticed when fluorinated additives were used to modify the PEDOT:PSS. Shading lights onto the effect of HTL fluorination, we show that the morphology of the polymer:PCBM blends remains surprisingly unaffected by the fluorinated HTL surface energy but that, instead, the OPVs are impacted not only by the HTL electronic properties (work function, dipole layer, open circuit voltage, charge transfer dynamic) but also by alteration of the complex refractive indices (photogeneration, short circuit current density, external quantum efficiencies, electro-optic modelling). Both mechanisms find their origin in fluorination induced charge transfers. This work points towards fluorination as a promising strategy toward combining both external quantum efficiency modulation and power conversion efficiency enhancement in OPVs. Charge transfers could also be used more broadly to tune the optical constants and electric field distribution, as well as to reduce interfacial charge recombinations within OPVs.

## Introduction

Organic and hybrid organic-inorganic optoelectronics are the subject of intensive research partially motivated by the potential to achieve low processing cost devices, for instance via roll-to-roll and inkjet printing processes, and by the promise to deliver exciting mechanical properties such as lightweight, flexibility and stretchability.[1-4] Organic materials for solution based processes range from low melting point semiconductors, suitable for liquid electronics, to large polymers, soluble in various solvents.[5-8] Ongoing molecular engineering efforts aim at combining solution properties with optimized energy levels and charge transport properties to design high performance devices including transistors,[9,10] light emitting diodes or cavities,[11,12] memories,[4,6] and solar cells.[2,13] Organic photovoltaic (OPV) devices in both single and multi-junction configuration have now reached in research environment the 10% efficiency seen as the breaking point announcing technology transfer and commercialization.[1,14-20] Among the suitable device structures, bulk hetero-junction (BHJ) solar cells are probably the most studied, with their specific morphology being seen as close to ideal: a) nanophase structuration to increase exciton dissociation, b) bi-continuous percolation network to allow efficient charge collection, c) donor (acceptor) rich phases near the hole (electron) transport layer to reduce charge recombination while increasing the charge selectivity of the electrodes. Various approaches have been developed to tune the phase separation and control the BHJ morphology. These include thermal and solvent vapour treatments, as well as the addition of small molecules or co-solvents, and the resulting structures have been investigated by ellipsometry, electron tomography, dynamic secondary ion mass spectroscopy, X-ray photoemission.[21-29]

Poly(3,4-ethylenedioxythiophene): poly(styrene sulfonate) (PEDOT:PSS, Fig. 1a1) is widely used as hole transport layers in optoelectronic devices. Sulfonate groups withdraw electrons from the PEDOT backbone and then transform the chain electronic state from neutral to polycationic. The PSS serve then the dual purpose of oxidizing PEDOT moieties and stabilizing in aqueous media the otherwise insoluble polymer. Stimulated by its unusual properties, PEDOT:PSS intensive study revealed that subtle changes in the molecule conformation could lead to dramatic alteration of its electronic properties. For this purpose a wide range of alcohol, acid, surfactant and polymer molecules have been used as co-solvents or as post-treatments of spin-coated films leading to conductivities up to 3000 S/cm and work functions ranging from 4.0 to 5.7 eV.[30-37] For instance, McCarthy et al. have recently reported on methanol and formic acid spray treatments of PEDOT:PSS films resulted in a 3 to 4 orders of magnitude improvement in sheet resistance values;[37] while Lipomi et al. have developed stretchable electrodes based on ultraviolet/ozone treated PDMS surfaces and the addition of Zonyl fluorosurfactant to a PEDOT:PSS solution.[34] The resulting materials have been used for light emitting diodes,[38-40] transistors,[41-44] heat flux sensors,[45-47] and solar cells.[48-53] In the latter case, PEDOT:PSS doping has been shown to increase photovoltaic efficiencies in both standard and inverted configurations.[53-60] In some cases, solvent treatments based on alcohol or acid derivatives have made possible the fabrication of ITO free devices with comparable *PCE*'s.[48,50,52,61,62]

Fluorinated materials have solution properties orthogonal to both water and oil derivatives. This can be of interest to create barriers or to control solubilization, surface properties, or even to complete chemical reactions in original environments.[63-66] Fluor is also widely used in push-pull molecular design due to its high electro-negativity.[67] A nonionic ethoxylated fluoro-surfactant was used to develop PDMS based stretchable electrodes for P3HT:$PC_{60}BM$ OPVs.[50] A similar nonionic material was blended with PEDOT:PSS to develop inverted solar cells with tunable efficiency and longer lifetime.[68] (Heptadecafluoro-1,1,2,2-tetra-hydrodecyl)triethoxysilane was spin-coated on-top of PEDOT:PSS layers subsequently annealed to create a silica based fluorinated spacer above which pentacene was deposited. This interface facilitated 3D single crystalline growth, and the development of interfacial dipole moments through the accumulation of negative charges which enhanced the built-in potential across the devices and resulted in increased open-circuit voltage, hole transport and device efficiency.[49] Germack et al. studied $SiO_2$, PEDOT:PSS and poly(thienothiophene):perfluorinated ionomer inter-faces with P3HT:$PC_{60}BM$ showing that whilst segregation at the buried interface near the HTL could be strongly affected by its surface energy, devices made with the latter two blends lead to OPVs with similar characteristics.[69-71]

However, little has been achieved to discriminate between the relative impacts of work function, conductivity, interfacial and optical property alterations including charge transfer. Herein, we gain original insight into this complex situation relying on a combination of experimental investigation including atomic force microscope, Kelvin probe, conductivity, device fabrication and characterisation, wetting, grazing incidence wide angle X-ray scattering, transient absorption and spectroscopic ellipsometry measurements, as well as devices electro-optical modelling. These, we show, provide a full picture of the system and are essential to gain insight into overlapping or complementary effects of fluorination agents (FAs). Two anionic fluorinated materials, i.e. perfluorinated ionomer (PFI, Fig. 1a2) and perfluorooctane sulfonic acid (FOS, Fig. 1a3) were selected to mix with PEDOT:PSS. Whilst FOS presents stronger environmental risks compared to PFI, the choice was motivated by their similar composition (fluorinated and sulfonic acid) allowing to compare the effects of fluorination with polymeric and small molecules on both PEDOT:PSS electronic properties and on device performances. We focused on PEDOT:PSS:FA solutions with weight ratio of 6:1:30, which was selected to observe noticeable effects without altering substantially the HTL conductivity while adding an excess of insulating materials. Fig. 1b/c shows the two photoactive blends, P3HT:$PC_{60}BM$ and PTB7:$PC_{70}BM$, which were used in this work. P3HT:$PC_{60}BM$ is certainly the most studied OPV model system; whereas, PTB7:$PC_{70}BM$ is a relatively newer model system. Interestingly, PTB7 is a low band-gap material, in which charge separation along its backbone appears to be enhanced by intramolecular charge separation associated with the alternating donor-acceptor groups delocalizing the excitons, lowering their binding energy and reducing charge carrier recombination.[72] *PCE*s up to 5.6 % have been achieved on semi-transparent substrate,[73] 8.7 % when it was blended with high mobility polymer,[18,74] 9.2 % with an inverted device structure,[75] and up to 10.1 % when dual side nanoimprint process were implemented.[19] A range of studies have been carried out on the effects of additives,[29,76-80] molecular weight,[81] processing,[82] and substrate,[83,84] on the BHJ morphology and efficiency. The combination of both photoactive blends is, however, convenient to strengthen and draw relevant comparisons.[83,85]

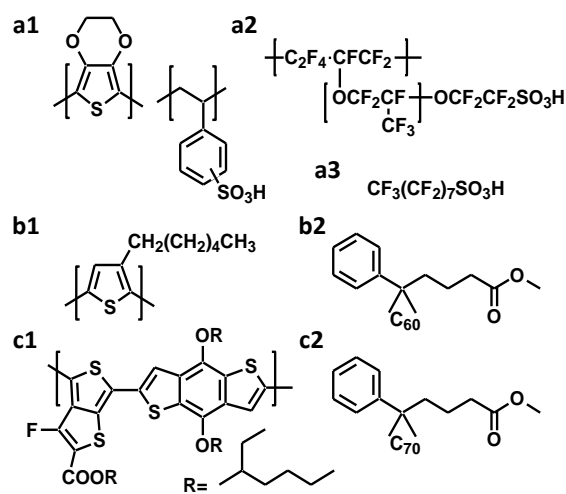

**Fig. 1.** Molecular structures of PEDOT: PSS **(a1)**, PFI **(a2)**, FOS **(a3)**, P3HT:$PC_{60}BM$ **(b1-2)**, PTB7: $PC_{70}BM$ **(c1-2)**.

In the present work, a stronger effect of the PEDOT:PSS fluorination is observed with PTB7 based devices than with P3HT OPVs, but both types of material display the same trend with HTL fluorination. We then show that the optoelectronic properties (work function, dipole layer, refractive index) of fluorinated PEDOT:PSS contribute greatly to the overall enhancement of the device efficiencies and their spectral modulations. In addition, and somewhat unexpectedly when compared with other systems, the variations of the wetting property and surface energy of the HTL layer appear to have no detectable influence on the photoactive blend morphology (crystallinity, orientation and composition profile), and then does not appear to be relevant to the variations of the performance of organic solar cells herein observed. In contrast, the electro-optical experimental and modelling results point towards the effect of charge transfers on optical constants to explain the OPVs efficiency spectral variation with HTL fluorination. This is an essential set of information which needs to be understood to take advantage of unusual features and compositions in BHJ devices.

## Results & Discussion

### PEDOT:PSS Fluorination

The first row of Fig. 2 presents typical atomic force microscopy (AFM) images. In solution, PEDOT:PSS forms micelles around a PEDOT crystalline core, which has been shown by scattering techniques to grow when PEDOT:PSS is subject to solvent treatments.[35] Once spin-coated, AFM images reveal bright and dark areas, which are commonly associated with PEDOT and PSS rich regions.[33,48,50-52,68] In the case of the mixed materials, PSS and the fluorination agents compete to stabilize the PEDOT polymeric chains. The morphology of the images is noticeably influenced by the presence of the fluorination agent leading to larger bright domains. This feature is more visible with the polymeric than with the surfactant fluorination (Fig. 2b1 vs c1). However, FOS based thin films reveals a finer substructure of smaller grains aggregated to form the bright regions. These variations would be consistent with conformational or aggregation changes of the polymer chains when co-solvents or fluorination agents are added. The finer substructure observed with FOS could results from its surfactant nature leading to a more effective distribution than with the polymer PFI. Also noticeable is the very small variation of the RMS and height between peaks and valley, which are only slightly larger in the fluorinated materials compared to pristine PEDOT-PSS thin films (Table 1, Fig. S1). These variations in the RMS and Peak to Peak values are not large enough to have any impact on device performances, which are presented in the following section.

Surface potential maps were obtained by Kelvin probe force (KPFM) microscopy and are displayed in the 2$^{nd}$ row of Fig. 2. Surface potential maps show three distinct average values, i.e. -220 mV, -720 mV and -250 mV for the pristine, PFI and FOS mixed PEDOT:PSS, respectively. Local variations of the surface potential are relatively small and of the order of ± 20 mV. As illustrated in Fig. 2b3 for the PFI based PEDOT:PSS sample, when the surface potential is overlaid onto the topography, the surface potential appears to be independent of the surface profile; Fig. S2 confirms this characteristic for all the samples.

From KPFM measurements, the local work function was deduced and its average values are presented in Table 1. Both fluorinated PEDOT:PSS films present a deeper work function than the pristine films. This is consistent with fluorinated materials located at the polymer-air interface due to their higher ionization potentials compared to alkyl chains.[39,86] For comparison purposes, the work functions of the thin films were also measured with a macroscopic Kelvin probe (Table 1).

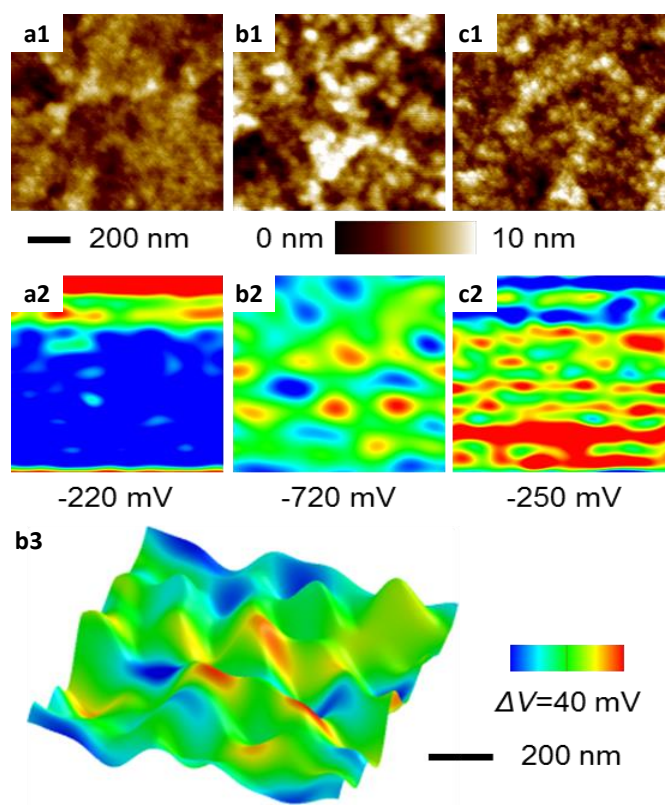

**Fig. 2.** PEDOT:PSS layers: mixed **(a)**, fluorinated with PFI **(b)** and FOS **(c)** topographic images obtained by atomic force microscopy **(1)**, surface potential maps obtained by Kelvin probe force microscopy **(2)**, surface potential on the topography for the PFI based PEDOT:PSS sample **(3)**.

**Table 1** PEDOT:PSS based thin film morphology as characterized by AFM in phase mode in Fig. 2: root mean square (*RMS*, ±0.1 nm) and peak-to-valley height ($h_{PtV}$), nanoscopic and macroscopic work functions ($W_{f\text{-KPFM}}$, $W_{f\text{-mKP}}$ ±0.02 eV).

| HTL Fluorination | x | PFI | FOS |
|---|---|---|---|
| *RMS* (nm) | 1.5 | 2.5 | 2.0 |
| $h_{PtV}$ (nm) | 11 ± 2 | 17 ± 1 | 14 ± 2 |
| $W_{f\text{-KPFM}}$ (eV) | 4.70 ± 0.02 | 5.40 ± 0.03 | 4.90 ± 0.03 |
| $W_{f\text{-mKP}}$ (eV) | 5.20 | 5.72 | 5.56 |

**Table 2** PEDOT:PSS based thin film conductivity ($\sigma$, mS/cm), just after annealing and followed by spin-coating pure solvent on-top before the electrode evaporation.

| HTL Fluorination | x | PFI | FOS |
| --- | --- | --- | --- |
| -- | 0.70 ± 0.05 | 0.55 ± 0.02 | 0.49 ± 0.03 |
| chlorobenzene | 0.69 ± 0.04 | 0.95 ± 0.09 | 0.55 ± 0.02 |
| 1,2-dichlorobenzene | 0.65 ± 0.07 | 0.83 ± 0.02 | 0.53 ± 0.02 |

The values differ slightly from those deduced from the KPFM measurements, i.e. between 5 and 12 %, however, they follow the same trend. The $W_{f\text{-}mKP}$ values of the fluorinated samples are much deeper than the value obtained with the pristine PEDOT:PSS film, and PFI-$W_{f\text{-}mKP}$ is also deeper than FOS-$W_{f\text{-}mKP}$. Consistent observations were made with the ionization potential deduced from UPS measurements (Fig. S3).

The main mechanisms of PEDOT:PSS conductivity are associated with the acid groups protonating the PEDOT and with charge hopping among the PEDOT polymer chains. The first column of Table 2 shows the conductivity of the spin-coated thin films, just after annealing as described in the experimental section. $\sigma_{HTL}$ is consistent with the literature values, however, it is noticeable that the PFI- and FOS- based samples present a slight decreased conductivity. The weak variation of PEDOT:PSS conductivity when fluorinated with PFI is consistent with the literature.[87]

Alongside of the already mentioned slight morphology alterations of the thin films observed by AFM (Fig. 2-row 1), both PFI and FOS should at least preserve the PEDOT protonation otherwise induced by PSS. The fluorinated nature of the fluorination agents is likely to favour the formation of larger and better connected PEDOT domains. However, the insulating PSS is then partially replaced by another insulating molecule, which overall increases the ratio of insulating to conductive materials in the HTL. Spin-coated PEDOT:PSS films are known to present an upper PSS rich phase,[39,40,88] hence the apparent decrease of the conductivity observed for the two fluorinated samples was assumed to result from a larger fraction of insulating material sitting at the top of the upper interface of both fluorinated PEDOT:PSS thin films. To assess this hypothesis, as well as the becoming of this insulating interfacial layer when an organic semiconductor thin film is spin-coated on the PEDOT:PSS, conductivity measurements were also completed after spin-coating a pure organic solvent on top of the annealed hole transport layer. Pure chlorobenzene and di-chlorobenzene were chosen to mimic the effect of spin-coating P3HT:$PC_{60}BM$ and PTB7:$PC_{70}BM$, respectively. The second and third rows of Table 2 show that the conductivity of pristine PEDOT:PSS films is unaltered by the organic solvent spin-coating process. FOS based thin films display a slight increase, ~ 10 %, of the films conductivity. PFI based PEDOT:PSS thin films show a more noticeable conductivity increase, i.e. ~ 50 % and ~ 70 % for dichloro-benzene and chlorobenzene, respectively. The apparent higher $\sigma$ values of the fluorinated films are likely due to a better connectivity between the electrodes and the underlying PEDOT.

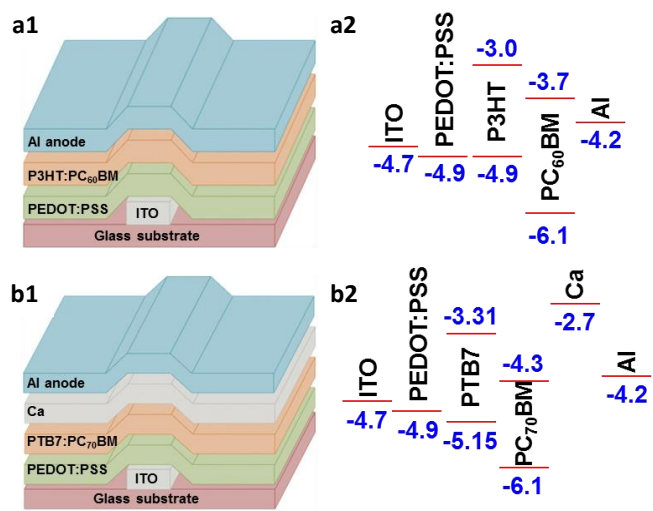

**Fig. 3.** Schematic of the devices **(1)** and the flat band energy diagram **(2)** associated with P3HT:$PC_{60}BM$ **(a)** and PTB7:$PC_{70}BM$ **(b)**.

The relative effect variation induced by spin-coating pure solvents on PEDOT:PSS fluorinated by small and large molecules could result from a different balance between the shear forces and the hydrophilic-fluorophilic-lipophilic character of each compound. Whilst, the observed enhancement is relevant, no dramatic change was observed and the obtained values remained within the range of conductivity usually associated with pristine PEDOS:PSS layers.

**Photovoltaic Devices and Time Resolved Spectroscopy**

The fluorinated-PEDOT:PSS thin films were used as hole transport layers in OPVs. The devices were based on two types of active semiconductor materials, P3HT:$PC_{60}BM$ (Fig. 1b) and PTB7:$PC_{70}BM$ (Fig. 1c). Fig. 3 presents both the device structures and their associated energy diagrams. Aluminium was evaporated on top of the P3HT:$PC_{60}BM$ active layer and the devices were characterized directly (Fig. 3b). In contrast, PTB7:$PC_{70}BM$ electron conduction layer relies on evaporation of both calcium and aluminium (Fig. 3a).[81,85] Consequently, these devices were encapsulated to address the high reactivity of the calcium layer under ambient conditions.

The normalized absorbance spectra characteristic of the semiconductor materials used in this investigation are presented in Fig. 4a/b1. PTB7 covers a wider spectral range than P3HT, and the relatively large fraction of fullerene derivatives can be noticed in the high energy part of the absorbance of the blend. Fig. 4a2 and b2 present the current density as a function of the applied voltage for both P3HT:$PC_{60}BM$ and PTB7:$PC_{70}BM$ devices, respectively. The characteristics of the devices are summarized in Table 3. Un-fluorinated PEDOT:PSS based devices display consistent efficiencies and characteristics as those reported in the literature. Incidentally, one could notice that chlorobenzene led to higher efficiency PTB7 solar cells when compared with OPVs prepared wih dichlorobenzene as a solvent. For both type of solar cell and both fluorination, the *PCE* of the device

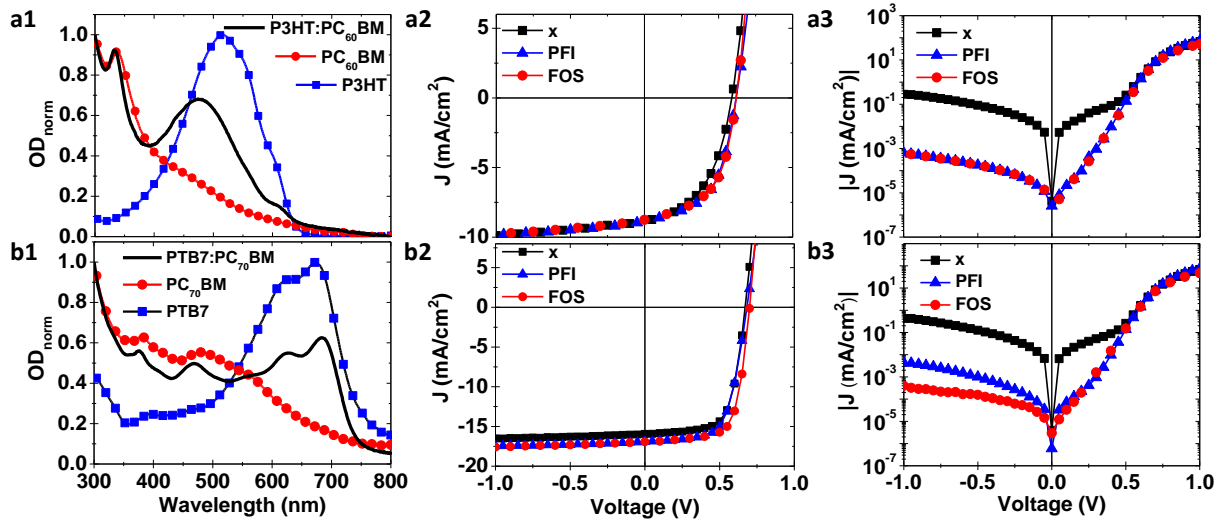

**Fig. 4.** P3HT:PC$_{60}$BM **(a)** and PTB7:PC$_{70}$BM **(b)**. Normalized absorbance spectra of films **(1)** made of the conjugated polymer (blue line and square), electron acceptor (red line and circles) and a blend of the two materials (black line). Solar cells prepared from these blends **(2)**, as described in the experimental section, and Dark J-V curves **(3)** with the PEDOT:PSS layer being pristine (■), mixed with PFI (▲) and with FOS (●).

**Table 3** Best (mean and standard deviation) device performances*: power conversion efficiency (*PCE*), open circuit voltage ($V_{OC}$), short circuit current density ($J_{SC}$), and fill factor (*FF*), external quantum efficiencies (*EQE**) at 450 nm and 0 V bias.

| HTL Fluorination | | x | PFI | FOS |
|---|---|---|---|---|
| P3HT:PC$_{60}$BM | PCE (%) | 2.53 (2.49±0.07) | 2.96 (2.91±0.05) | 2.91 (2.85±0.05) |
| | $V_{OC}$ (V) | 0.59 (0.59±0.01) | 0.62 (0.64±0.01) | 0.63 (0.64±0.02) |
| | FF (%) | 48.5 (49.7±2.6) | 54.0 (52.9±0.7) | 53.3 (53.0±0.9) |
| | $J_{SC}$ (mA/cm$^2$) | 8.84 (8.48±0.32) | 8.86 (8.60±0.17) | 8.63 (8.47±0.12) |
| | EQE* (%) | 55.4 | 57.7 | 59.8 |
| PTB7:PC$_{70}$BM | PCE (%) | 7.18 (6.77±0.26) | 7.49 (7.30±0.14) | 8.26 (8.12±0.15) |
| | $V_{OC}$ (V) | 0.67 (0.68±0.01) | 0.68 (0.68±0.01) | 0.70 (0.70±0.01) |
| | FF (%) | 67.2 (63.5±2.3) | 65.0 (64.3±1.3) | 69.3 (68.8±1.7) |
| | $J_{SC}$ (mA/cm$^2$) | 15.94 (15.76±0.36) | 16.89 (16.71±0.23) | 16.94 (16.78±0.17) |
| | EQE* (%) | 72.5 | 77.1 | 79.4 |

* The OPV area was 0.08 cm$^2$ and the statistic was established with at least 7 devices.

shown to increase when compared with the performance obtained with the pristine hole conduction layer. The highest PCE is obtained with PEDOT:PSS:FOS for PTB7:PC$_{70}$BM, while there is only a marginal difference between FOS and PFI based OPVs in the case of P3HT:PC$_{60}$BM. For the two types of device, the fluorination of PEDOT:PSS translates into an increase of $V_{OC}$, with a maximum variation of 45 mV and 33 mV for P3HT and PTB7 based devices, respectively. This is consistent with the work function alteration induced by the HTL fluorination (Table 1) and suggests a lower amount of recombination in the device. This is thought to occur by an alteration of the band bending and the internal electric field at the HTL/heterojunction interface which prevent electrons from recombining at the hole-extracting electrode.[89-91] The fill factor is more sensitive and the least understood but depends on the charge accumulation at the electrodes, i.e. balanced charge extraction, and molecular charge recombination, which depending upon the system can be geminate or non-geminate.[92-95] P3HT:PC$_{60}$BM fill factors increase with the fluorination of the HTL, with a maximum difference of ~ 5.5 %, i.e. ~11.3 % relative variation. This is consistent with the $V_{OC}$ variation and could be associated with a decrease of the resistive losses. The average FF values of PTB7:PC$_{70}$BM devices displays a similar trend, however the very small FF difference relative variation compared to pristine HTL device and the relatively large standard deviation prevents us to draw any reliable conclusion from the FF variation of the low band-gap BHJ devices. The $J_{SC}$ data display a relatively low variation, ~ 0.2 mA/cm$^2$, when the fluorination agents are used to fabricate the P3HT:PC$_{60}$BM devices. However, a larger increase, ~ 0.9 mA/cm$^2$, occurs in the case of PTB7:PC$_{70}$BM, which suggests that in this later case a larger amount of charges being photogenerated upon fluorination of the HTL. Incidentally, we note that the devices fabricated with either of the additive in the HTL systematically present lower standard variations of the PCE, FF and $J_{SC}$. This is observed for both polymer:PCBM blends, as well as for both fluorinated additives, and as a consequence is attributed to the fluorinated nature of the additives.

The dark J-V curves are presented in Fig. 4a3 and b3. The classical "diode curve" shape is observed for the un-fluorinated PEDOT:PSS based devices.[85] From 1.0 to 0.5 V the curves display the usual bell shape; at 0.5 V bias, a change of slope is noticeable with a much slower decay, which changes plateau around 10$^{-2}$ mA/cm$^2$ for very small positive external bias. In reversed bias, the current density increases slowly and continuously to reach a maximum absolute value, which is slightly larger than of 10$^{-2}$ mA/cm$^2$. Whilst the fluorinated PEDOT:PSS devices present exactly the same bell shape between 1.0 and 0.5 V, the slope of the continuous decay is unaltered below 0.5 V, with the leakage current density reaching a minimum absolute value close to 10$^{-4}$ mA/cm$^2$ for small and positive external bias. In reversed bias, the current density increases slowly and continuously to reach maximum absolute values under -1 V bias of the order of 10$^{-3}$ mA/cm$^2$. The P3HT:PC$_{60}$BM dark J-V curves of both PFI and FOS mixed with PEDOT:PSS overlap almost perfectly. PTB7:PC$_{70}$BM devices present the same trend, even though for reversed bias those with PFI based HTL display slightly larger current density absolute values than with FOS, which is slightly larger than 10$^{-2}$ mA/cm$^2$. In other systems, such leakage current decreases were associated with higher PCBM segregation at the anodes and consequently enhanced contact selectivity. Fig. 5a1 and a2 present the external quantum efficiencies (*EQE*) of the devices. The *EQE*s of pristine PEDOT:PSS devices are consistent with the literature for each type of semiconductors,[21,96] with PTB7:PC$_{70}$BM covering a much broader spectral range than P3HT:PC$_{60}$BM, as already mentioned when discussing their respective absorption spectra presented in Fig. 4. The *EQE* of the latter blend peaks around 500 nm and displays a kink at high energy. PTB7 devices fabricated from dichlorobenzene have a rather flat *EQE* response across 350-750 nm spectral range. Both P3HT:PC$_{60}$BM and PTB7:PC$_{70}$BM show slight increase of the *EQE* with PEDOT:PSS fluorination. For illustration purposes, Table 3 presents the 450 nm *EQE* values for each device, these *EQE* measurements are independent of *PCE*, $V_{oc}$ and *FF* but do relate to $J_{sc}$. In the case of P3HT:PC$_{60}$BM, the same small variation and trend is observed for both the integral of *EQE* and $J_{sc}$ data, with up to 8 % increase with PEDOT:PSS fluorination. For the PTB7 solar cells, fluorination induces a consistent 6 to 9 % relative variation of both integrated *EQE* and $J_{sc}$, leading also to higher values than in P3HT based OPVs.

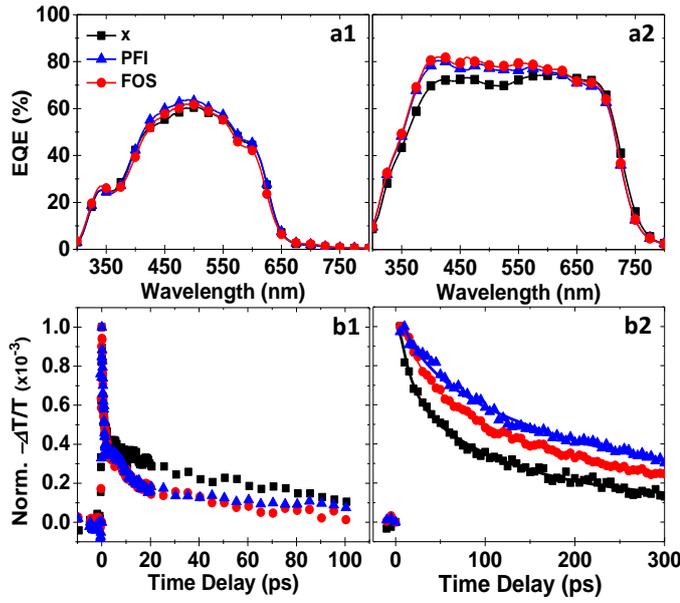

**Fig. 5.** Device external quantum efficiencies (a) and transient absorption measurements (b) for P3HT:PC$_{60}$BM (1) and PTB7: PC$_{70}$BM (2) thin films on-top of PEDOT:PSS HTL pristine (■), mixed with PFI (▲) and with FOS (●).

**Table 4** Transient absorption fit parameters and calculated errors for both P3HT:PC$_{60}$BM ($\lambda_{probe}$ = 640 nm) and PTB7:PC$_{70}$BM ($\lambda_{probe}$ = 810 nm) deposited on top of PEDOT:PSS based thin films and photoexcited at room temperature with a $\lambda_{pump}$ = 400 nm pulsed beam.

| | Fluorination | x | PFI | FOS |
|---|---|---|---|---|
| P3HT: PC$_{60}$BM | $A_1$ | 0.59 ± 0.01 | 0.64 ± 0.01 | 0.64 ± 0.01 |
| | $\tau_{CR1}$ (ps) | 0.63 ± 0.05 | 0.75 ± 0.05 | 0.84 ± 0.05 |
| | $A_2$ | 0.41 ± 0.01 | 0.36 ± 0.01 | 0.36 ± 0.01 |
| | $\tau_{CR2}$ (ps) | 74 ± 3 | 33 ± 3 | 29 ± 2 |
| PTB7: PC$_{70}$BM | $A_1$ | 0.47 ± 0.02 | 0.33 ± 0.03 | 0.42 ± 0.02 |
| | $\tau_{CR1}$ (ps) | 16.9 ± 1.4 | 36.6 ± 4.2 | 41.3 ± 2.5 |
| | $A_2$ | 0.53 ± 0.02 | 0.67 ± 0.05 | 0.58 ± 0.03 |
| | $\tau_{CR2}$ (ps) | 211 ± 9 | 352 ± 22 | 333.4 ± 17 |

Consequently, two enhancement mechanisms appear to be involved upon HTL fluorination, *i*) an increased charge extraction preventing their accumulation, and *ii*) an enhanced charge photogeneration. The relative contribution of these mechanisms differs whether P3HT:PC$_{60}$BM or PTB7:PC$_{70}$BM devices are considered. The first mechanism could be associated with a faster sweep of the charges out of the blend, which consequently would not allow their recombination. Using deeper work function can lead to a more favourable band bending at the contact interface. This alters the energy alignment within the devices and, as described by the Integer Charge Transfer model, creates a strong surface dipole, which assists the charge extraction from the interface before they can recombine.[89-91] This interpretation is supported by the transient absorption measurements presented in Fig. 5b. Upon photoexcitation, ultrafast charge separations occur and form a charge transfer (CT) state, which leads to either geminate recombination or the separation of charges into the donor-acceptor bi-continuous network. These free carriers can eventually recombine as non-geminate pairs. For the two types of blends herein investigated, we note that a) charge separation occurs over too short of a time scale to be accessed with our setup, however, b) the charge recombination (CR) kinetics appear to be clearly altered by the fluorination of the HTL. The P3HT:PC$_{60}$BM (Fig. 5b1) and PTB7: PC$_{70}$BM (Fig. 5b2) samples were probed at 640 and 810 nm, respectively. The use of these wavelengths implies that in the first blend, it is the P3HT cation which are probed,[97,98] whereas the kinetics of the PC$_{70}$BM anions are monitored in the second blend.[99,100] The CR kinetics were fitted with at least two characteristic times, which values and weights are summarised in Table 4. P3HT:PC$_{60}$BM thin films present characteristic times of disappearance of the P3HT cationic species in the picosecond, $\tau_{CR1}$, and tens of picosecond, $\tau_{CR2}$, ranges. $\tau_{CR1}$ could contain a contribution from geminate recombination, which will likely be less affected by PEDOT:PSS fluorination. In any case, the experimental and fit uncertainties prevent discussing the $\tau_{CR1}$ 18 and 33 % variations with PEDOT:PSS fluorination. More interestingly, by fluorination the PEDOT:PSS layer, the longer characteristic time $\tau_{CR2}$ is drastically decreased by more 55 and 60 % for PFI and FOS agents, respectively. $\tau_{CR2}$ is understood as containing both non-geminate recombination and hole transfer from the P3HT cations to the PEDOT:PSS layer. An accelerated sweep of the holes from the blends is consistent with the larger work functions reported in Table 1 and with dipole formation associated to charge transfers at the interface of P3HT and fluorinated PEDOT:PSS. The negatively charged fluorinated PEDOT:PSS then accelerates the extraction of the photo-induced holes, leading to shorter $\tau_{CR2}$ values, and this is consistent with both the $V_{OC}$ increase and the enhanced PCE of the devices fabricated with fluorinated HTL (Table 3). In a similar manner, PTB7:PC$_{70}$BM systems display a bi-exponential kinetic, with characteristic times in the 10s and 100s of picosecond range, i.e. slower than the P3HT:PC$_{60}$BM blends. A strong effect of the HTL fluorination on the charge transfer state kinetics is clear from Fig. 5 and Table 4. The PC$_{70}$BM anion disappearance characteristic times are increased by more than 117 and 145 % for $\tau_{CR1}$, as well as 67 and 59 % for $\tau_{CR2}$, upon PFI and FOS agents, respectively. This substantial effect is again consistent with observations made by monitoring the P3HT cation dynamics. Probing the PC$_{70}$BM acceptors reveals that when more holes are swept out from the blends to the HTL, the recombination probability of the PC$_{70}$BM anions is decreased due to a lower density of holes in the blends. This results in fewer charge recombinations at the interface of the hole-extracting electrode, and is consistent with the increased $V_{OC}$ when the HTL is fluorinated as presented in Table 3. This more efficient charge extraction can be associated with the lower deprotonation energy of the sulfonate resulting from a higher fluorination of the PEDOT and with a higher built-in potential in the device. This is equivalently described through the higher dipole moments of the deprotonated fluorinated sulfonic acid materials compared with alkyl chains derivatives.[86-88,101,102]

However, the spectral features of the *EQE* spectra require further careful considerations as the HTL fluorination does not result in a homogeneous increase of the *EQE* reference spectra. On the contrary, the *EQE* ratios of the fluorinated and unfluorinated devices vary with the wavelength (Fig. 5a). P3HT:PC$_{60}$BM OPVs display a maximum increase of 4 % of the EQE narrowly located between 400 and 500 nm. The PTB7:PC$_{70}$BM devices present an *EQE* increase with fluorination of up to 12 % spreading between 300 and 600 nm, with stronger variations near 400 and 500 nm. This spectral variation cannot be explained by a change of transmission of the fluorinated PEDOT:PSS layer, which shows a negligible, i.e. < 1 %, variation around its maximum (Fig. S7). In addition, the two blends do not show the same *EQE* spectral variation with PEDOT:PSS fluorination. Several hypotheses can then be made including a) the fluorination altering the HTL surface energy, which in return impact on a1) the material packing and distribution within the blends, and a2) the thickness of the device layers, b) HTL:blend interface charge transfers altering the optical constants of the materials involved.
Any of these hypotheses could be responsible of the *EQE* spectral variations observed in Fig. 5a as a1) would result in a change of absorption coefficients, a2) and b) would impact the electric field distribution within the devices.

**Interfacial Effects and Modelling**

The variation of the surface energy with the PEDOT:PSS fluori-nation was assessed by contact angle measurements, which are presented in Table S3 and Fig. S8. The fluorinated agents lead to a lower surface energy when compared with un-fluorinated PEDOT:PSS. This could then provide a PCBM depleted region at the interface of the hole extraction layer,[69-71] which could contribute at preventing charge recombination at the surface of the hole-extracting electrode, and would be consistent with the increased $V_{OC}$. To explain the *EQE* wavelength dependence, a variation of surface energy and a local change of P3HT and PTB7 concentrations relative to the PCBM electron acceptor moieties would need to be associated with an alteration of the crystalline order of the P3HT or PTB7 at least near the HTL interface. The impact of surface energy on organic semiconductor crystallinity has been documented in the context of field effect transistors,[10,49,103-105] even though most studies have focused on pentacene. In the present OPV context, an alteration of the HTL surface energy could result from a change of phase separation and blend morphology analogous to the effect obtained when the solvent is slowly evaporated from the photoactive layer or when an additive is used to control the morphology of bulk heterojunction.[15,21-29,106,107] These have been suggested in P3HT:PCBM BHJ deposited on PEDOT:PSS with surface energy ranging from 50 to 70 mN/m, even though we note that ellipsometry measurements were associated with device properties and not with any in-depth structural studies.[71] P3HT:PC$_{60}$BM presenting the weakest *EQE* spectral variation, we first focused on PTB7:PC$_{70}$BM to compare the devices absorbance with and without fluorination of the HTL. Compared to the PEDOT:PSS based devices, the fluorinated OPVs do display an increased absorbance, which maximum is of the order of 6 % and which within the experimental precision would be consistent with the *EQE* spectral variation (Fig. S5b and S6). In addition, correlations between *EQE* and crystalline structure have been pointed at in systems including P3HT,[21,49,108] and PBTTPD.[25] However, we note that P3HT:PC$_{60}$BM does not present any obvious change of absorption, not even around 620 nm, which is associated to this polymer crystallization. Consequently and despite the contact angle data, the absorption results do not unambiguously support the hypothesis that the morphology, i.e. the crystallization, of the active layer is altered by the fluorination of the PEDOT:PSS layer. To resolve this issue, we completed a careful GIWAXS investigation, which is reported in section 4.4 of the SI. The insertion of the PCBM acceptor moieties in the conjugated polymers was shown to induce slight structural variations such as thickness, correlation length and alignment of lamellae; however, these features were shown for both P3HT:PC$_{60}$BM and PTB7:PC$_{70}$BM not to be affected by the fluorination of the HTL. This consequently excludes hypothesis a1).
The thickness of each layer was measured both by dektak and spectroscopic ellipsometry. The techniques gave comparable thicknesses for the blends, while the deviation was smaller than 10 % for the ellipsometry and within the experimental uncertainty for the dektak. Then, there is no strong thickness alteration, which could explain the *EQE* spectral variation, and this rules out the hypothesis a2). It is also noticed that the abs-ence of significant thickness variation excludes any mobility change, which otherwise could have altered the *PCE*.[109,110]

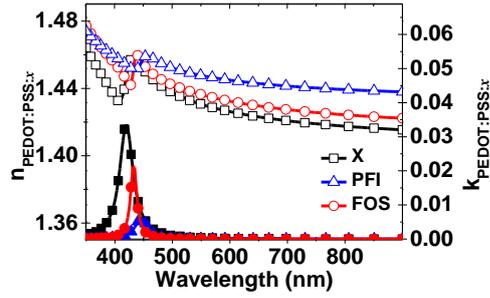

**Fig. 6.** PEDOT:PSS refractive indices and extinction coefficients: pristine (■), mixed with PFI (▲) and with FOS (●).

To explore the last hypothesis, it is essential to consider the role of the optical constants and electric field within the devices. As described with more details in the section SI-4.5.2, fluorinated PEDOT:PSS refractive index and extinction coefficient were shown to present a red shift and an amplitude decrease of their main features when compared with pristine PEDOT:PSS thin films (Fig. 6). They are related to the excess of sulfonic acid groups brought in by the PFI and FOS agents along with the electro-negativity of the fluorinated moieties. This is equivalent to negative Burstein shifts induced on the apparent band gap when extra charges are added, for instance through doping,[111-114] and consistent with the variation of $n$ and $k$ observed for various conjugated molecules when in oxidized or reduced states.[115-117] Noticeably, this trend was preserved when the PEDOT:PSS based thin films were covered by the conjugated polymer:PCBM blends (Fig. S17). Considering that the number of excited state at the position, $z$, within a device layer depends on the energy which is locally absorbed, the device efficiency is then directly proportional to the time average of the energy dissipated per second, $Q$, which it-self is associated with the interferences between incident and reflected light. Under normal incidence, it can be expressed as:[118,119]

$$Q_j(z) = \tfrac{1}{2} c \varepsilon_0 \alpha_j n_j |E_j(z)|^2 \quad (1)$$
$$\alpha_j = \tfrac{4\pi k_j}{\lambda} \quad (2)$$

with $j$, the layer under consideration, the speed of light, $c$, the permittivity of the free space, $\varepsilon_0$, the complex refractive index, $\tilde{n} = n + ik$, the real part of refractive index, $n$, the extinction coefficient, $k$, the attenuation or absorption coefficient, $\alpha$, the electric field, $E$, and the wavelength, $\lambda$, of the incident light. While preserving, for sake of simplicity, we define $\tilde{Q}$ as

$$\tilde{Q}_j(z) = \tfrac{Q_j(z)}{2\pi c \varepsilon_0} = \tfrac{k_j n_j}{\lambda} |E_j(z)|^2 \quad (3)$$

Fig. 7 presents the distribution inside the devices of the modulus squared of the optical electric field, $|E|^2$, which was calculated as a function of the incident wavelength. Fig. 7a and b are associated with P3HT:PC$_{60}$BM and PTB7:PC$_{70}$BM, respectively. The dimensionless parameter $z/L$ is used for convenience purposes to materialise the different material layers forming the devices. As expected, $|E|^2$ is not monotonous with both incident wavelength and position within the devices. Noticeably, the $|E|^2$ maximum value is higher in the case of PTB7:PC$_{70}$BM than with P3HT:PC$_{60}$BM. Similar patterns were obtained with fluorinated PEDOT:PSS as shown in Fig. S19. The vertical dashed line in Fig. 7 is a cross-section of $|E|^2$ at 500 nm illumination.

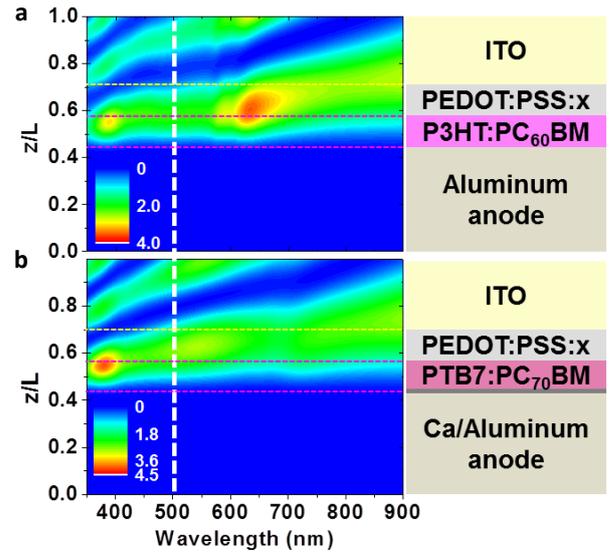

**Fig. 7.** Calculated distribution of the modulus squared of the optical electric field, $|E|^2$, as a function of the incident wavelength inside a photovoltaic device made of P3HT:PC$_{60}$BM **(a)** and PTB7:PC$_{70}$BM **(b)** spincoated on top of pristine PEDOT:PSS. The horizontal dotted lines stand for the material distribution as labelled on the right hand-side.

A similar oscillating behaviour of $|E|^2$ is observed in Fig. 8a1 and b1 for the three HTLs. Systematically, $|E|^2$ tails off in the metal anode and its maximum occurs in the PEDOT:PSS and blend based layers. As described in eq. 3, $\tilde{Q}$ was calculated by taking into account *n* and *k* values of each layer, and is shown in Fig. 8-2. Regardless of the fluorination and blend, $\tilde{Q}$ is equal to zero and very small in the metal anode and HTL, respectively. In Fig. 8a2, we notice that with P3HT:PC$_{60}$BM, the maximum of $\tilde{Q}$ is marginally larger with the fluorinated HTLs than with the pristine PEDOT:PSS. However, in the case of PTB7:PC$_{70}$BM this trend is much more pronounced as the $\tilde{Q}$ values of the fluorinated HTLs remain larger than those of the pristine HTL all across the photoactive layers. The inserts of Fig. 8-2 focus on these photoactive blend layers and matches the z/L value corresponding to the blend to HTL interfaces. The overall variation of $\tilde{Q}$ between the devices was quantified by integrating the $\tilde{Q}$ values as a function of *z/L*. The areas under the $\tilde{Q}$ curves are summarized in Table 5. When compared with pristine HTL, the increase of $\tilde{Q}$ is of the order of 7 % in the case of P3HT:PC$_{60}$BM. PTB7:PC$_{70}$BM based devices show an increase of $\tilde{Q}$ of about 18 and 35 % for PFI and FOS based HTL, respectively. We confirmed with PTB7:PC$_{70}$BM that such an increase could not be induced by a change of the PEDOT:PSS layer thickness (Fig. S20). We also note that comparing Table 3 and 5, the variations of $\tilde{Q}$ and $J_{SC}$ are consistent with one another, weak for P3HT:PC$_{60}$BM but pronounced for PTB7:PC$_{70}$BM.

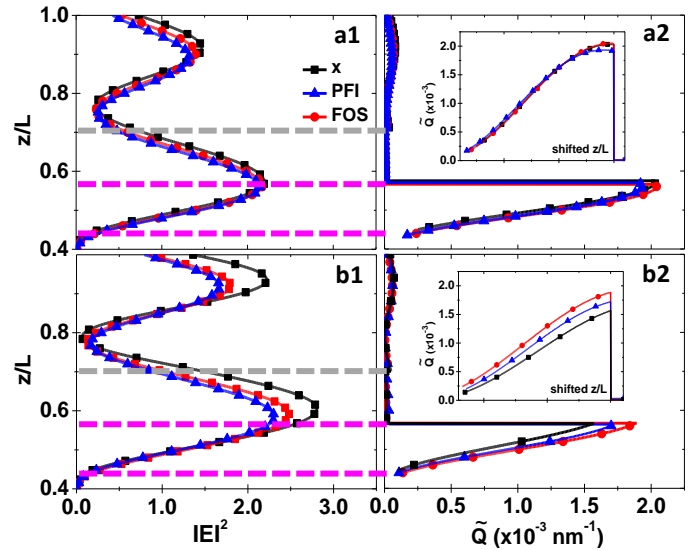

**Fig. 8.** Cross-section at 500 nm illumination of $|E|^2$ **(1)** across the photovoltaic devices made of P3HT:PC$_{60}$BM **(a)** and PTB7:PC$_{70}$BM **(b)** spincoated on top of pristine PEDOT:PSS. The horizontal dashed lines correspond to the interfaces between each layer. Calculated $\tilde{Q}$ values **(2)**. Insert: $\tilde{Q}$ values in the polymer:blend layers with z/L shifted to overlap the blend:anode interface. PEDOT:PSS HTL layer: pristine (■), mixed with PFI (▲) and with FOS (●).

**Table 5** Components of the complex refractive index of the polymer:PCBM blend at 500 nm, with *n*, the real part, and *k*, the extinction coefficient, used to calculate $\tilde{Q}_{int.}$, the integral of $\tilde{Q}$ across the device photoactive layer.

| HTL Fluorination | | x | PFI | FOS |
|---|---|---|---|---|
| P3HT: PC$_{60}$BM | *n* | 1.796 | 1.819 | 1.817 |
| | *k* | 0.258 | 0.248 | 0.255 |
| | $\tilde{Q}_{int.}$ (10$^{-4}$) | 1.54 | 1.65 | 1.65 |
| PTB7: PC$_{70}$BM | *n* | 1.573 | 1.665 | 1.596 |
| | *k* | 0.196 | 0.232 | 0.248 |
| | $\tilde{Q}_{int.}$ (10$^{-4}$) | 0.99 | 1.17 | 1.44 |

Whilst the electro-optical analysis remains partially subject to the models and fits used to analyse the spectroscopic ellipsometry data, it does highlights the potential of fluorination agent to alter the optical constants of OPV layers. It is important to keep in mind that these charge transfers are independent of any photo-excitation, and instead induced by the sulfonic acids groups pending at the apex of the fluorination agent in the HTL. The variations of the HTL work functions and refractive index differ by the fact that the former applies to photo-induced charges by sweeping them out of the photoactive blend, while the latter is spectrally resolved, as illustrated by the transient absorption, the *EQE* curves and the electro-optical modelling. The latter contribution has usually been neither investigated, nor taken advantage off.

## Conclusion

The fluorination of the hole transport layer by fluorinated molecules have been characterized by near-field microscopy showing that the morphology of the thin films was not drastically altered by the fluorination agent. In contrast, Kelvin probe measurements have revealed a large impact on the substrate work function, which remained very homogeneous even at the nanometer scale. The conductivity of the hole transport layers were shown to decrease slightly when the fluorinated sulfonic acid surfactant and the sulfonic ionomer were used as fluorination agent. The film conductivity was altered further when fresh solvent was spin-coated on top of the HTL. Two different photoactive polymer:PCBM blends were used to fabricate solar cells on top of un-fluorinated and fluorinated PEDOT:PSS thin films. As for both small surfactants and large polymeric macromolecules, the device efficiencies were increased, any specific influence of the fluorination agent molecular structure and conformation could be excluded. GIWAXS data showed that the interfacial surface energy and wetting properties had no effect on the morphology, crystallinity and donor-acceptor distribution of the photoactive polymer:PCBM blends. Regardless of the conjugated polymer: PCBM blend, when using the fluorinated additives, the device characteristics were shown to be systematically more reproducible from one device to another. The increased power conversion efficiency of the devices based on fluorinated PEDOT:PSS layer was shown to be solely related to the electronic and optical properties of the fluorinated hole transport layer, through an alteration of not only its work function but also its refractive index. Two distinct mechanisms lead to *i*) an increased charge extraction ($V_{OC}$, $\tau_{CR}$) and *ii*) an enhanced charge photogene-ration ($J_{SC}$, *n-k*, $\tilde{Q}$), which relative contributions vary with the conjugated polymer:PCBM blend. These cannot be separated as they occur simultaneously and share the same origin, i.e. charge transfers induced by the fluorination agents.

Overall, the present results shed lights onto the optoelectronic effects on BHJ OPVs of using fluorinated agent in HTL preparations and are likely to be applicable to electrode interlayers. They also suggest that whilst it can be of interest to alter the electrode with a dipole layer to aim at increasing the interfacial charge extraction, fluorination can also be used more broadly to tune the optical constants and electric field distribution within the devices. Herein, such a strategy is shown to lead to a power conversion efficiency increase of up to 15 % along with a noticeable change of the device external quantum efficiencies in the UV-visible range. Impacts will likely vary in strength and spectral range from one fluorination agent, polymeric electrode, or photoactive blend, to another, so that this work paves the way toward a broad range of materials to be systematically explored as we demonstrated that this fluorination strategy is an important and widely applicable parameter to control increase OPV characteristics.

## Supplementary Information:

Experimental Section and further details about phase and surface potential near-field measurements, transmittance, wetting, GIWAXS, ellipsometry, transient absorption and calculations of internal electric field and absorption.

## Acknowledgements

PA thanks the Canon Foundation in Europe for supporting his visits to the RIKEN through a personal Fellowship, and the OSC to access the facilities. KJL, HK, JHK, ESK, JWW, DWK and PA were supported by funding of the Ministry of Science, ICT & Future Planning, Korea (201000453, 2015001948, 2014M3A6B3063706). Part of this work has been carried out in the framework of CNRS International Associated Laboratory "Functional nanostructures: morphology, nanoelectronics and ultrafast optics" (LIA NANOFUNC), France. TA would like to acknowledge funding from the Japanese Society for the Promotion of Science via a JSPS KAKENHI grant (no. 22350084). JCR and CA would like to acknowledge financial support from the Regional Innovation Strategy Support Program -Kumamoto Area on Organic electronics collaboration - Ministry of Education, Culture, Sports, Science and Technology (MEXT). The authors thank Pohang Accelerator Laboratory (PAL) in South Korea for giving us the opportunity to perform the GIWAXS measurements in the frame of the proposal number "2014-1st-9A-015". The authors are grateful to MEST and POSTECH for supporting these experiments, to Drs. Tae Joo Shin and Hyungju Ahn for adjustments and assistance, as well as to other staff members from 9A U-SAXS beamline for further assistance.

## Notes and references

1. L. Dou, J. You, Z. Hong, Z. Xu, G. Li, R. A. Street and Y. Yang, *Adv. Mater.*, 2013, **25**, 6642.
2. F. Cicoira and C. Santato, *Organic Electronics: Emerging Concepts and Technologies*, Wiley, 2013.
3. M. Kuik, G. J. Wetzelaer, H. T. Nicolai, N. I. Craciun, D. M. De Leeuw and P. W. Blom, *Adv. Mater.*, 2014, **26**, 512.
4. R. H. Kim, H. J. Kim, I. Bae, S. K. Hwang, D. B. Velusamy, S. M. Cho, K. Takaishi, T. Muto, D. Hashizume, M. Uchiyama, P. Andre, F. Mathevet, B. Heinrich, T. Aoyama, D. E. Kim, H. Lee, J. C. Ribierre and C. Park, *Nat. Comm.*, 2014, **5**, 3583.
5. S. Hirata, K. Kubota, H. H. Jung, O. Hirata, K. Goushi, M. Yahiro and C. Adachi, *Adv. Mater.*, 2011, **23**, 889.
6. J. C. Ribierre, T. Aoyama, T. Muto and P. André, *Org. Electron.*, 2011, **12**, 1800.
7. A. Mishra and P. Bauerle, *Ang. Chem. Int. Ed.*, 2012, **51**, 2020.
8. S. Watanabe, Y. Fukuchi, M. Fukasawa, T. Sassa, M. Uchiyama, T. Yamashita, M. Matsumoto and T. Aoyama, *Langmuir*, 2012, **28**, 10305.
9. K. Muhieddine, M. Ullah, B. N. Pal, P. Burn and E. B. Namdas, *Adv. Mater.*, 2014, **26**, 6410.
10. J. C. Ribierre, L. Zhao, S. Furukawa, T. Kikitsu, D. Inoue, A. Muranaka, K. Takaishi, T. Muto, S. Matsumoto, D. Hashizume, M. Uchiyama, P. Andre, C. Adachi and T. Aoyama, *Chem. Comm.*, 2015, **51**, 5836.
11. D. Ballarini, M. De Giorgi, S. Gambino, G. Lerario, M. Mazzeo, A. Genco, G. Accorsi, C. Giansante, S. Colella, S. D'Agostino, P. Cazzato, D. Sanvitto and G. Gigli, *Adv. Opt. Mater.*, 2014, **2**, 1076.
12. H. Kaji, H. Suzuki, T. Fukushima, K. Shizu, K. Suzuki, S. Kubo, T. Komino, H. Oiwa, F. Suzuki, A. Wakamiya, Y. Murata and C. Adachi, *Nat. Commun.*, 2015, **6**, 8476.
13. J. Min, H. Zhang, T. Stubhan, Y. N. Luponosov, M. Kraft, S. A. Ponomarenko, T. Ameri, U. Scherf and C. J. Brabec, *J. Mater. Chem. A*, 2013, **1**, 11306.
14. R. A. Janssen and J. Nelson, *Adv. Mater.*, 2013, **25**, 1847.
15. Y. Liu, C. C. Chen, Z. Hong, J. Gao, Y. Michael Yang, H. Zhou, L. Dou, G. Li and Y. Yang, *Sci. Rep.*, 2013, **3**, 3356.
16. A. R. b. M. Yusoff, D. Kim, H. P. Kim, F. K. Shneider, W. J. da Silva and J. Jang, *Energy Environ. Sci.*, 2015, **8**, 303.
17. N. Li and C. J. Brabec, *Energy Environ. Sci.*, 2015, **8**, 2902.
18. S. Liu, P. You, J. Li, J. Li, C.-S. Lee, B. S. Ong, C. Surya and F. Yan, *Energy Environ. Sci.*, 2015, **8**, 1463.
19. J. D. Chen, C. Cui, Y. Q. Li, L. Zhou, Q. D. Ou, C. Li, Y. Li and J. X. Tang, *Adv. Mater.*, 2015, **27**, 1035.
20. W. Yu, L. Huang, D. Yang, P. Fu, L. Zhou, J. Zhang and C. Li, *J. Mater. Chem. A*, 2015, **3**, 10660.
21. M. Campoy-Quiles, T. Ferenczi, T. Agostinelli, P. G. Etchegoin, Y. Kim, T. D. Anthopoulos, P. N. Stavrinou, D. D. C. Bradley and J. Nelson, *Nat. Mater.*, 2008, **7**, 158.
22. Y. Yao, J. Hou, Z. Xu, G. Li and Y. Yang, *Adv. Funct. Mater.*, 2008, **18**, 1783.
23. J. Jo, S.-I. Na, S.-S. Kim, T.-W. Lee, Y. Chung, S.-J. Kang, D. Vak and D.-Y. Kim, *Adv. Funct. Mater.*, 2009, **19**, 2398.
24. S. S. van-Bavel, E. Sourty, G. de With and J. Loos, *Nano Lett.*, 2009, **9**, 507.
25. M. S. Su, C. Y. Kuo, M. C. Yuan, U. S. Jeng, C. J. Su and K. H. Wei, *Adv. Mater.*, 2011, **23**, 3315.
26. E. J. W. Crossland, K. Rahimi, G. Reiter, U. Steiner and S. Ludwigs, *Adv. Funct. Mater.*, 2011, **21**, 518.
27. Y. Vaynzof, D. Kabra, L. Zhao, L. L. Chua, U. Steiner and R. H. Friend, *ACS Nano*, 2011, **5**, 329.
28. M. Kim, J.-H. Kim, H. H. Choi, J. H. Park, S. B. Jo, M. Sim, J. S. Kim, H. Jinnai, Y. D. Park and K. Cho, *Adv. Energy Mater.*, 2014, **4**, 1300612.
29. B. A. Collins, Z. Li, J. R. Tumbleston, E. Gann, C. R. McNeill and H. Ade, *Adv. Energy Mater.*, 2013, **3**, 65.
30. J. Huang, P. F. Miller, J. S. Wilson, A. J. de Mello, J. C. de Mello and D. D. C. Bradley, *Adv. Funct. Mater.*, 2005, **15**, 290.
31. J. Ouyang, Q. Xu, C.-W. Chu, Y. Yang, G. Li and J. Shinar, *Polymer*, 2004, **45**, 8443.
32. A. M. Nardes, R. A. J. Janssen and M. Kemerink, *Adv. Funct. Mater.*, 2008, **18**, 865.
33. J.-H. Huang, D. Kekuda, C.-W. Chu and K.-C. Ho, *J. Mater. Chem.*, 2009, **19**, 3704.
34. D. J. Lipomi, J. A. Lee, M. Vosgueritchian, B. C. K. Tee, J. A. Bolander and Z. A. Bao, *Chem. Mater.*, 2012, **24**, 373.
35. T. Takano, H. Masunaga, A. Fujiwara, H. Okuzaki and T. Sasaki, *Macromolecules*, 2012, **45**, 3859.
36. Y. Xia, K. Sun and J. Ouyang, *Adv. Mater.*, 2012, **24**, 2436.
37. J. E. McCarthy, C. A. Hanley, L. J. Brennan, V. G. Lambertini and Y. K. Gun'ko, *J. Mater. Chem. C*, 2014, **2**, 764.
38. X. Crispin, F. L. E. Jakobsson, A. Crispin, P. C. M. Grim, P. Andersson, A. Volodin, C. van Haesendonck, M. Van der Auweraer, W. R. Salaneck and M. Berggren, *Chem. Mater.*, 2006, **18**, 4354.
39. T. W. Lee, Y. Chung, O. Kwon and J. J. Park, *Adv. Funct. Mater.*, 2007, **17**, 390.
40. T.-W. Lee and Y. Chung, *Adv. Funct. Mater.*, 2008, **18**, 2246.
41. D. W. Li and L. J. Guo, *J. Phys. D-Appl. Phys.*, 2008, **41**, 105115.
42. V. Sholin, S. A. Carter, R. A. Street and A. C. Arias, *Appl. Phys. Lett.*, 2008, **92**, 063307.
43. I. Pang, H. Kim, S. Kim, K. Jeong, H. S. Jung, C.-J. Yu, H. Soh and J. Lee, *Org. Electron.*, 2010, **11**, 338.
44. H. Tang, P. Lin, H. L. W. Chan and F. Yan, *Biosens. Bioelectron.*, 2011, **26**, 4559.
45. J. J. Luo, D. Billep, T. Waechtler, T. Otto, M. Toader, O. Gordan, E. Sheremet, J. Martin, M. Hietschold, D. R. T. Zahnd and T. Gessner, *J. Mater. Chem. A*, 2013, **1**, 7576.
46. N. Massonnet, A. Carella, O. Jaudouin, Patrice Rannou, G. Laval, C. Celle and J.-P. Simonato, *J. Mater. Chem. C*, 2014, **2**, 1278.
47. T. Park, C. Park, B. Kim, H. Shin and E. Kim, *Energy Environ. Sci.*, 2013, **6**, 788.
48. Y. H. Kim, C. Sachse, M. L. Machala, C. May, L. Müller-Meskamp and K. Leo, *Adv. Funct. Mater.*, 2011, **21**, 1076.
49. H. C. Han, C. A. Tseng, C. Y. Du, A. Ganguly, C. W. Chong, S. B. Wang, C. F. Lin, S. H. Chang, C. C. Su, J. H. Lee, K. H. Chen and L. C. Chen, *J. Mater. Chem.*, 2012, **22**, 22899.
50. M. Vosgueritchian, D. J. Lipomi and Z. A. Bao, *Adv. Funct. Mater.*, 2012, **22**, 421.

51. D. Alemu, H.-Y. Wei, K.-C. Ho and C.-W. Chu, *Energy Environ. Sci.*, 2012, **5**, 9662.
52. W. Zhang, B. Zhao, Z. He, X. Zhao, H. Wang, S. Yang, H. Wu and Y. Cao, *Energy Environ. Sci.*, 2013, **6**, 1956.
53. J.-H. Kim, S.-Y. Huh and S. Seo, *Jpn. J. Appl. Phys.*, 2014, **53**, 04ER03.
54. Z. Y. Hu, J. J. Zhang, Z. H. Hao and Y. Zhao, *Sol. Energy Mater. Sol. Cells*, 2011, **95**, 2763.
55. J. Ouyang, C. W. Chu, F. C. Chen, Q. Xu and Y. Yang, *Adv. Funct. Mater.*, 2005, **15**, 203.
56. A. M. Nardes, M. Kemerink, M. M. de Kok, E. Vinken, K. Maturova and R. A. J. Janssen, *Org. Electron.*, 2008, **9**, 727.
57. K. M. Kim, K. W. Lee, A. Moujoud, S. H. Oh and H. J. Kim, *Electrochem. Solid State Lett.*, 2010, **13**, H447.
58. S. Park, S. J. Tark and D. Kim, *Curr. Appl. Phys.*, 2011, **11**, 1299.
59. C.-J. Huang, K.-L. Chen, Y.-J. Tsao, D.-W. Chou, W.-R. Chen and T.-H. Meen, *Synt. Met.*, 2013, **164**, 38.
60. Z. Hu, J. Zhang and Y. Zhu, *Renew. Energy*, 2014, **62**, 100.
61. M. A. Rahman, A. Rahim, M. Maniruzzaman, K. Yang, C. Lee, H. Nam, H. Soh and J. Lee, *Sol. Energy Mater. Sol. Cells*, 2011, **95**, 3573.
62. M. Song, H.-J. Kim, C. S. Kim, J.-H. Jeong, C. Cho, J.-Y. Lee, S.-H. Jin, D.-G. Choi and D.-H. Kim, *J. Mater. Chem. A*, 2015, **3**, 65.
63. J. M. DeSimone and W. Tumas, *Green chemistry using liquid and supercritical carbon dioxide*, Oxford University Press, Oxford, 2003.
64. P. Lacroix-Desmazes, P. André, J. M. DeSimone, A.-V. Ruzette and B. Boutevin, *J. Polym. Sci. Pol. Chem.*, 2004, **42**, 3537.
65. J. Guo, P. André, M. Adam, S. Panyukov, M. Rubinstein and J. M. DeSimone, *Macromolecules*, 2006, **39**, 3427.
66. P. Andre, P. Lacroix-Desmazes, D. K. Taylor and B. Boutevin, *J. Supercrit. Fluids*, 2006, **37**, 263.
67. H. S. Nalwa and S. Miyata, *Nonlinear optics of organic molecules and polymers*, CRC Press, Boca Raton, Fla., 1997.
68. F. J. Lim, K. Ananthanarayanan, J. Luther and G. W. Ho, *J. Mater. Chem.*, 2012, **22**, 25057.
69. D. S. Germack, C. K. Chan, R. J. Kline, D. A. Fischer, D. J. Gundlach, M. F. Toney, L. J. Richter and D. M. DeLongchamp, *Macromolecules*, 2010, **43**, 3828.
70. D. S. Germack, C. K. Chan, B. H. Hamadani, L. J. Richter, D. A. Fischer, D. J. Gundlach and D. M. DeLongchamp, *Appl. Phys. Lett.*, 2009, **94**, 233303.
71. P. G. Karagiannidis, N. Kalfagiannis, D. Georgiou, A. Laskarakis, N. A. Hastas, C. Pitsalidis and S. Logothetidis, *J. Mater. Chem.*, 2012, **22**, 14624.
72. J. M. Szarko, B. S. Rolczynski, S. J. Lou, T. Xu, J. Strzalka, T. J. Marks, L. Yu and L. X. Chen, *Adv. Funct. Mater.*, 2014, **24**, 10.
73. R. Betancur, P. Romero-Gomez, A. Martinez-Otero, X. Elias, M. Maymo and J. Martorell, *Nat. Phot.*, 2013, **7**, 995.
74. T. Goh, J.-S. Huang, B. Bartolome, M. Y. Sfeir, M. Vaisman, M. L. Lee and A. D. Taylor, *J. Mater. Chem. A*, 2015, **3**, 18611.
75. Z. He, C. Zhong, S. Su, M. Xu, H. Wu and Y. Cao, *Nat. Phot.*, 2012, **6**, 593.
76. S. J. Lou, J. M. Szarko, T. Xu, L. Yu, T. J. Marks and L. X. Chen, *J. Am. Chem. Soc.*, 2011, **133**, 20661.
77. P. A. Cox, D. A. Waldow, T. J. Dupper, S. Jesse and D. S. Ginger, *ACS Nano*, 2013, **7**, 10405.
78. Z. Li, G. Lakhwani, N. C. Greenham and C. R. McNeill, *J. Appl. Phys.*, 2013, **114**, 034502.
79. S. Guo, E. M. Herzig, A. Naumann, G. Tainter, J. Perlich and P. Muller-Buschbaum, *J. Phys. Chem. B*, 2014, **118**, 344.
80. C.-G. Wu, C.-H. Chiang and H.-C. Han, *J. Mater. Chem. A*, 2014, **2**, 5295.
81. C. Liu, K. Wang, X. Hu, Y. Yang, C. H. Hsu, W. Zhang, S. Xiao, X. Gong and Y. Cao, *ACS Appl. Mater. Interfaces*, 2013, **5**, 12163.
82. I.-J. No, P.-K. Shin, S. Kannappan, P. Kumar and S. Ochiai, *Appl. Phys. Res.*, 2012, **4**, 83.
83. N. Zhou, D. B. Buchholz, G. Zhu, X. Yu, H. Lin, A. Facchetti, T. J. Marks and R. P. Chang, *Adv. Mater.*, 2014, **26**, 1098.
84. T. Yanagidate, S. Fujii, M. Ohzeki, Y. Yanagi, Y. Arai, T. Okukawa, A. Yoshida, H. Kataura and Y. Nishioka, *Jpn. J. Appl. Phys.*, 2014, **53**, 02BE05.
85. A. Guerrero, N. F. Montcada, J. Ajuria, I. Etxebarria, R. Pacios, G. Garcia-Belmonte and E. Palomares, *J. Mater. Chem. A*, 2013, **1**, 12345.
86. J. S. Kim, J. H. Park, J. H. Lee, J. Jo, D.-Y. Kim and K. Cho, *Appl. Phys. Lett.*, 2007, **91**, 112111.
87. Y. W. Zhu, T. Song, F. T. Zhang, S. T. Lee and B. Q. Sun, *Appl. Phys. Lett.*, 2013, **102**, 113504.
88. M.-R. Choi, T.-H. Han, K.-G. Lim, S.-H. Woo, D. H. Huh and T.-W. Lee, *Ang. Chem. Int. Ed.*, 2011, **123**, 6398.
89. K. M. Knesting, H. Ju, C. W. Schlenker, A. J. Giordano, A. Garcia, O. N. L. Smith, D. C. Olson, S. R. Marder and D. S. Ginger, *J. Phys. Chem. Lett.*, 2013, **4**, 4038.
90. J. C. Blakesley and D. Neher, *Phys. Rev. B*, 2011, **84**, 075210.
91. I. Hancox, L. A. Rochford, D. Clare, M. Walker, J. J. Mudd, P. Sullivan, S. Schumann, C. F. McConville and T. S. Jones, *J. Phys. Chem. C*, 2013, **117**, 49.
92. B. Qi and J. Wang, *Phys. Chem. Chem. Phys.*, 2013, **15**, 8972.
93. G. F. Dibb, F. C. Jamieson, A. Maurano, J. Nelson and J. R. Durrant, *J. Phys. Chem. Lett.*, 2013, **4**, 803.
94. L. Wu, H. Zang, Y.-C. Hsiao, X. Zhang and B. Hu, *Appl. Phys. Lett.*, 2014, **104**, 153903.
95. D. Bartesaghi, C. Perez Idel, J. Kniepert, S. Roland, M. Turbiez, D. Neher and L. J. Koster, *Nat. Commun.*, 2015, **6**, 7083.
96. Y. Liang, Z. Xu, J. Xia, S. T. Tsai, Y. Wu, G. Li, C. Ray and L. Yu, *Adv. Mater.*, 2010, **22**, E135.
97. I. W. Hwang, D. Moses and A. J. Heeger, *J. Phys. Chem. C*, 2008, **112**, 4350.
98. G. Grancini, D. Polli, D. Fazzi, J. Cabanillas-Gonzalez, G. Cerullo and G. Lanzani, *J. Phys. Chem. Lett.*, 2011, **2**, 1099.
99. T. Nojiri, M. M. Alam, H. Konami, A. Watanabe and O. Ito, *J. Phys. Chem. A*, 1997, **101**, 7943.
100. K. Yonezawa, M. Ito, H. Kamioka, T. Yasuda, L. Han and Y. Moritomo, *Adv. Opt. Tech.*, 2012, **2012**, 316045.
101. T.-W. Lee, O. Kwon, M.-G. Kim, S. H. Park, J. Chung, S. Y. Kim, Y. Chung, J.-Y. Park, E. Han, D. H. Huh, J.-J. Park and L. Pu, *Appl. Phys. Lett.*, 2005, **87**, 231106.
102. S. A. Mauger, J. Li, Ö. T. Özmen, A. Y. Yang, S. Friedrich, M. D. Rail, L. A. Berben and A. J. Moulé, *J. Mater. Chem. C*, 2014, **2**, 115.
103. W. Shao, H. Dong, L. Jiang and W. Hu, *Chem. Sci.*, 2011, **2**, 590.
104. C. Liu, Y. Xu and Y.-Y. Noh, *Mater. Today*, 2015, **18**, 79.
105. D. J. Gundlach, J. E. Royer, S. K. Park, S. Subramanian, O. D. Jurchescu, B. H. Hamadani, A. J. Moad, R. J. Kline, L. C. Teague, O. Kirillov, C. A. Richter, J. G. Kushmerick, L. J. Richter, S. R. Parkin, T. N. Jackson and J. E. Anthony, *Nat. Mater.*, 2008, **7**, 216.
106. K. Sun, Z. Xiao, E. Hanssen, M. F. G. Klein, H. H. Dam, M. Pfaff, D. Gerthsen, W. W. H. Wong and D. J. Jones, *J. Mater. Chem. A*, 2014, **2**, 9048.
107. J. R. Tumbleston, B. A. Collins, L. Yang, A. C. Stuart, E. Gann, W. Ma, W. You and H. Ade, *Nat. Phot.*, 2014, **8**, 385.
108. Z. Xu, L.-M. Chen, G. Yang, C.-H. Huang, J. Hou, Y. Wu, G. Li, C.-S. Hsu and Y. Yang, *Adv. Funct. Mater.*, 2009, **19**, 1227.
109. C. H. To, A. Ng, Q. Dong, A. B. Djurisic, J. A. Zapien, W. K. Chan and C. Surya, *ACS Appl. Mater. Interfaces*, 2015, **7**, 13198.
110. B. Huang, E. Glynos, B. Frieberg, H. Yang and P. F. Green, *ACS Appl. Mater. Interfaces*, 2012, **4**, 5204.
111. E. Burstein, *Phys. Rev.*, 1954, **93**, 632.
112. B. R. Bennett, R. A. Soref and J. A. Del Alamo, *IEEE J. Quantum Electron.*, 1990, **26**, 113.
113. G. Lakhwani, R. F. H. Roijmans, A. J. Kronemeijer, J. Gilot, R. A. J. Janssen and S. C. J. Meskers, *J. Phys. Chem. C*, 2010, **114**, 14804.
114. A. R. S. Kandada, S. Guarnera, F. Tassone, G. Lanzani and A. Petrozza, *Adv. Funct. Mater.*, 2014, **24**, 3094.
115. D. Rae Kim, W. Cha and W.-k. Paik, *Synt. Met.*, 1997, **84**, 759.
116. D. Kim, Duckhwan Lee and W.-k. Paik, *Bull. Korean Chem. Soc.*, 1996, **17**, 707.
117. M. T. Giacomini, L. M. M. de Souza and E. A. Ticianelli, *Surf. Scie.*, 1998, **409**, 465.
118. L. A. A. Pettersson, L. S. Roman and O. Inganäs, *J. Appl. Phys.*, 1999, **86**, 487.
119. D. W. Sievers, V. Shrotriya and Y. Yang, *J. Appl. Phys.*, 2006, **100**, 114509.